\documentstyle[11pt]{article}
\makeatletter
\def\@maketitle{\newpage   
 \null
 \vspace*{-1\headsep}      
 \vspace*{-1\headheight}
 \begin{flushright}{\normalsize	   
   \@date}
 \end{flushright}
 \vskip \headheight   
 \begin{center}		   
   {\LARGE \@title \par}
   \vskip 1.5em
   {\large
     \lineskip .5em
     \begin{tabular}[t]{c}\@author
     \end{tabular}\par}
   \vskip 1em
 \end{center}
 \par}
\makeatother
\renewcommand {\topfraction}{0.8}
\renewcommand {\textfraction}{0.2}
\renewcommand {\floatpagefraction}{0.7}

\begin{document}

\title{Asymptotic behavior of A + B $\rightarrow$ inert for particles with a
	drift\thanks{Supported in part by NSF Mathematical Sciences
	Postdoctoral Research Fellowship DMS 90-07206}}
\author{S. A. Janowsky\\Department of Mathematics\\University of Texas\\
                Austin, TX 78712\\janowsky@math.utexas.edu}
\date{January 1994}
\maketitle
\renewcommand{\baselinestretch}{1.7}\large\normalsize
\begin{abstract}
We consider the asymptotic behavior of the (one dimensional) two-species
annihilation reaction A + B $\rightarrow$ 0, where both species have a
uniform drift in the same direction and like species have a hard core
exclusion.  Extensive numerical simulations show that starting with an
initially random distribution of A's and B's at equal concentration the
density decays like $t^{-1/3}$ for long times.  This process is thus in
a different universality class from the cases without drift or with drift
in different directions for the different species.

\end{abstract}

\renewcommand{\baselinestretch}{1.85}\large\normalsize

The irreversible two species annihilation reaction A + B $\rightarrow$ 0
has been studied for quite some time as an example of a reaction
diffusion process where fluctuations are important, so that the density
decays more slowly than one would predict from the mean field rate
equations (at least in low dimension)~\cite{TW,KR1,KR2,BL,LR,RL}.  The
standard picture~\cite{TW} is that in a region of size $L$ it takes a
time on the order of $L^2$ for all the particles to react, since they
must diffuse around the region in order to annihilate.  The remaining
density will be proportional to the intial excess of either type A or
type B particles in this region, which is proportional to the square
root of the volume.  Thus at time $t$ one expects that the concentration
$c(t)$ will behave as
\begin{equation}
c(t) \sim \left[c(0)L^d\right]^{1/2} \Big/ L^d \sim
	\left[c(0)t^{d/2}\right]^{1/2} \Big/ t^{d/2}
	=c(0)^{1/2} t^{-d/4},
\end{equation}
for dimension $d\le4$, {\em i.e.}\ $c(0)^{1/2} t^{-1/4}$ in dimension
one, which has been verified rigorously~\cite{BL}.
In comparison the mean field result is $c(t) = [t + 1/c(0)]^{-1}$.

In order for this picture to be valid, it is necessary that the
distribution of particles be actually determined by diffusion.  The
addition of a drift field can invalidate this assumption.

That this happens when the two species drift in opposite directions is
not unexpected.  If the two species have a relative velocity $v$, then
using the same reasoning as above, one determines the concentration to
be~\cite{EF,BF}
\begin{equation}
c(t) \sim \left[c(0)vt^{(d+1)/2}\right]^{1/2} \Big/ \left[vt^{(d+1)/2}\right]
	=[c(0)v]^{1/2} t^{-(d+1)/4}
\end{equation}
for dimension $d\le3$, or $[c(0)v]^{1/2} t^{-1/2}$ in one dimension.

What is perhaps unexpected is that if the two species have a drift in
the {\em same} direction, the result may still be different from the
expected $t^{-d/4}$.  This is because even after subtracting off the
average motion, one does not necessarily recover diffusive behavior.

In order to proceed, we must specify the model further.  To keep matters
simple we consider a strong drift---particles move either to the right
or not at all.  We consider two types of particles moving on the one
dimensional lattice (with periodic boundary conditions).  At each
(micro-)time step, we randomly pick a site.  If that site is occupied,
we attempt to move the particle to the right.  If the site on the right
is unoccupied, the jump succeeds.  If the site is occupied by a particle
of the same species, the jump fails.  If the site is occupied by a
particle of the opposite species, both particles annihilate.

We now see why the diffusive picture might be invalid.  If we just
considered one species of particles, the model described above is the
Asymmetric Simple Exclusion Process (ASEP), and the long time behavior
is not governed by a linear diffusion equation.  Instead one must
consider a nonlinear stochastic equation such as the noisy Burgers
equation~\cite{Spohn},
\begin{equation}
\frac{\partial \rho}{\partial t} + \rho\frac{\partial \rho}{\partial x}
= \nu\frac{\partial^2\rho}{\partial x^2} + \frac{\partial\xi}{\partial x},
\label{Burgers}
\end{equation}
where $\rho$ is a rescaled density, $\nu$ is the viscosity (representing
the lattice spacing) and $\xi$ is a random noise term, {\em e.g.} white
noise where the covariance is
\begin{equation}
\langle \xi(x,t)\xi(x',t') \rangle = 2\nu\delta(x-x')\delta(t-t').
\end{equation}
Thus the exclusion rule, even for very low densities, can play a crucial
role in the dynamics of the annihilation process, contrary to the
popular view.

\begin{figure}
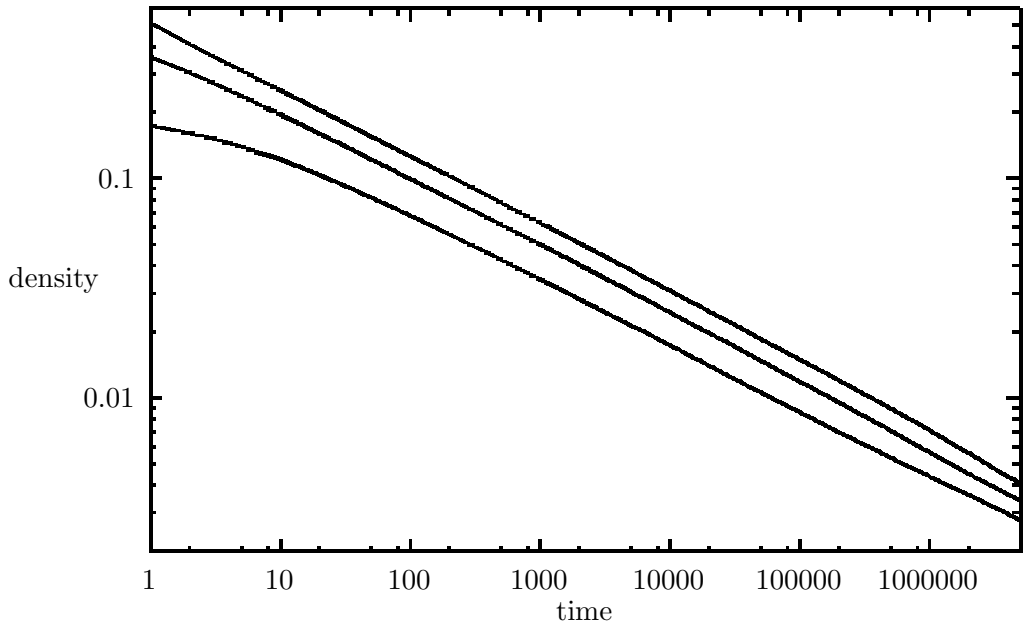

\begin{center}
\setlength{\unitlength}{0.2409pt}
\ifx\plotpoint\undefined\newsavebox{\plotpoint}\fi

\caption{Log-log plot of density vs.\ time for system of size $4\times10^6$.
Initial densities are 0.9, 0.5 and 0.2.\label{dens-time}}
\end{center}
\end{figure}

We studied the time evolution of the density for a variety of system
sizes and initial densities (always taking equal concentration of A and
B particles).  We considered systems up to size $4\times10^6$
for up to $5\times10^6$ time steps.  Some example runs are shown in
figure~\ref{dens-time}.
We see that qualitatively the asymptotic behavior seems to be
independent of the initial density, and that there appears to be a power
law decay of the density with exponent approximately equal to $-1/3$.

\begin{figure}
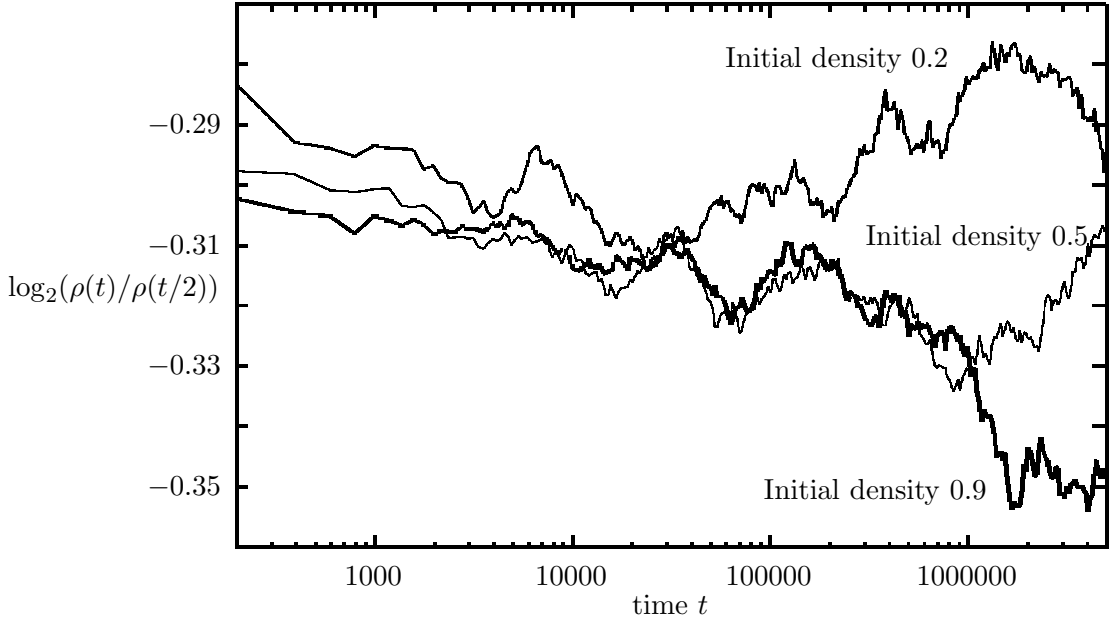

\begin{center}
\setlength{\unitlength}{0.2409pt}
\ifx\plotpoint\undefined\newsavebox{\plotpoint}\fi

\caption{Semilog plot of $\log_2 (\rho(t)/\rho(t/2))$ vs.\ time for
system of size $4\times10^6$.  Initial densities are 0.9, 0.5 and
0.2.\label{slope-time}}
\end{center}
\end{figure}

\begin{figure}
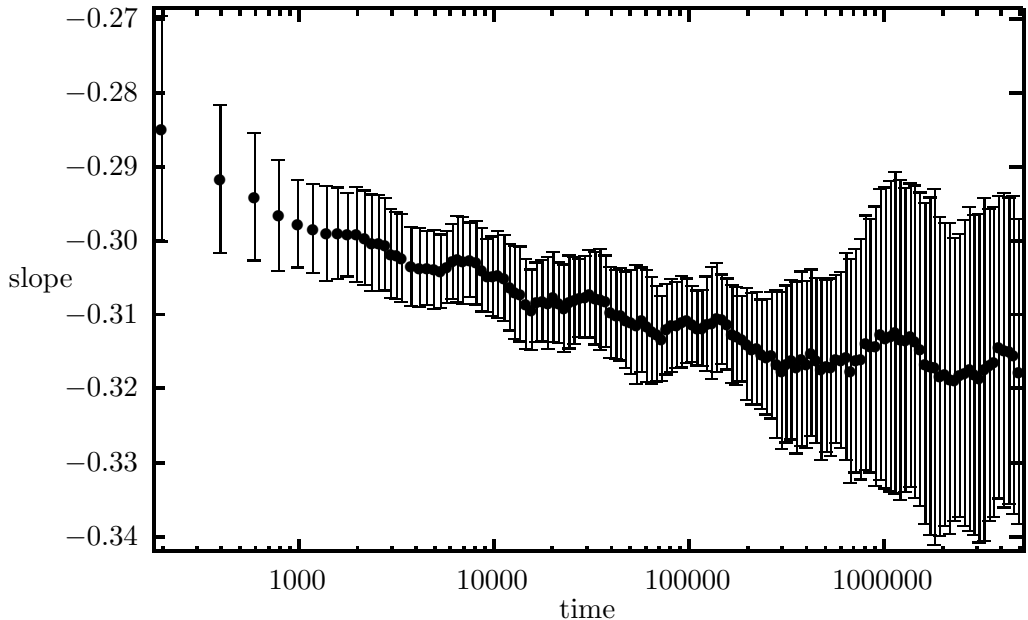

\begin{center}
\setlength{\unitlength}{0.2409pt}
\ifx\plotpoint\undefined\newsavebox{\plotpoint}\fi

\caption{Semilog plot of average slope vs.\ time for system of size
$4\times10^6$.\label{avg-slope-time}}
\end{center}
\end{figure}

To get a better view of the slope of these curves we examined the
behavior of $\rho(t)/\rho(t/2)$ as a function of time.  Specifically,
$\log_2 [\rho(t)/\rho(t/2)]$ should give the slope of the line ({\em
i.e.}\ the exponent of the power law behavior).  Data for
the same three runs as in figure~\ref{dens-time} are plotted in
figure~\ref{slope-time}.
This confirms that the decay behaves similarly for the different initial
densities, although the noise is now much more apparent.  To reduce the
effect of this, we averaged the slopes illustrated in
figure~\ref{slope-time} over 11 different initial densities.  The
resulting data is reproduced in figure~\ref{avg-slope-time}.  The
errorbars represent the standard deviation of the different slopes.

The errorbars at early times are primarily systematic in origin.  They
represent the fact that the decay rates are initially different for
different initial densities---the particle motion is diffusive in nature
until the particles have had a chance to interact with each other, so
that we see $t^{-1/4}$ behavior at early times for low density.  On the
other hand the errors at large time are primarily random errors;
fluctuations in the initial density are significantly amplified when the
density gets small.


Extrapolating the data of figure~\ref{avg-slope-time} to infinite time
is problematic.  It is not clear whether the effective exponent will
continue to decrease; all one can say with confidence is that it is less
than $-0.31$.  There are arguments, however, for believing that the
answer should be $-1/3$.

The exponent $-1/4$ for the diffusive case arises because the relevant
diffusive length scale at time $t$ is $t^{1/2}$.  On the other hand, in
the noisy Burgers equation (\ref{Burgers}) and thus in the one
dimensional ASEP, there is superdiffusive behavior, so that fluctuations
spread more quickly and the length scale is larger.  Thus the relevant
length scale is $t^{2/3}$~\cite{vBKS,GS}.  Thus one might expect the
concentration to go as
\begin{equation}
c(t) \sim L^{1/2} / L \sim
	\left[t^{2/3}\right]^{1/2} \Big/ t^{2/3}
	= t^{-1/3},
\end{equation}
certainly consistent with what is observed.  Of course this argument is
neglects the fact that only isolated clusters of one species undergo
asymmetric simple exclusion dynamics, but one expects that clusters each
species do spend most of their time reasonably isolated from each
other~\cite{LR,RL}, so that this analysis should be applicable.

\renewcommand{\baselinestretch}{1.17}\large\normalsize
\subsection*{Acknowledgments}
I would like to thank V. Belitsky for helpful discussions.

\end{document}